\documentclass{aa}
\usepackage{graphicx}
\usepackage{txfonts}
\begin{document}
\title{First VLBI mapping of circumstellar $^{29}$SiO maser emission}
   
   \author{R. Soria-Ruiz\inst{1} \and F. Colomer\inst{2} \and J.
     Alcolea\inst{1} \and V. Bujarrabal\inst{2} \and J.-F.
     Desmurs\inst{1} \and K.B. Marvel\inst{3}
    }

   \offprints{{R.\,Soria-Ruiz}\\ \email{r.soria@oan.es}}

   \institute{Observatorio Astron\'omico Nacional, Alfonso XII 3,
     E-28014 Madrid, Spain \and Observatorio Astron\'omico Nacional,
     Apartado 112, E-28803 Alcal\'a de Henares, Spain \and American
     Astronomical Society, 2000 Florida Avenue, NW Suite 400,
     Washington, DC 20009-1231, USA
    }

   \date{Received 15 December 2004 / Accepted 1 February 2005}
   
   \abstract{ We report the first VLBI map of the $v$=0 $J$=1--0 maser
     line of $^{29}$SiO in the long-period variable star IRC\,+10011.
     We have found that this maser emission is composed of multiple
     spots distributed in an incomplete ring, suggesting that this
     maser is also amplified tangentially, as already proposed in
     other SiO circumstellar masers. We present also VLBI maps for the
     7\,mm $^{28}$SiO $v$=1 and 2 $J$=1--0 and the 3\,mm $v$=1
     $J$=2--1 lines. The $^{29}$SiO masing region appears to be
     located in a layer in between the $^{28}$SiO $v$=1 $J$=1--0 and
     $^{28}$SiO $v$=1 $J$=2--1 lines. In addition, we confirm that the
     86 GHz maser $v$=1 $J$=2--1 forms in an outer region of the
     circumstellar envelope compared to the other $^{28}$SiO masers
     studied.  Finally, we discuss the possible implications of the
     observational results on the SiO maser pumping theory.

      }

   \authorrunning{R. Soria-Ruiz et al.}
   \titlerunning{mm-VLBI observations of $^{29}$SiO in IRC+10011}

   \maketitle
%

 \section{Introduction}
 
 SiO maser emission is found in the innermost shells of the
 circumstellar envelopes of Long-Period Variable (LPV) stars, evolving
 along the Asymptotic Giant Branch.  To date, three different
 isotopomers of the SiO molecule are known to exhibit maser emission
 towards these objects: $^{28}$SiO, $^{29}$SiO and $^{30}$SiO, with
 relative abundances of [$^{28}$SiO]/[$^{29}$SiO]$\sim$\,20 and
 [$^{28}$SiO]/[$^{30}$SiO]$\sim$\,30 respectively.
 
 Strong $^{28}$SiO maser lines have been detected, either by
 single-dish or interferometry techniques, in hundreds of evolved
 stars, from the $v$=0 to the $v$=4 levels and up to the rotational
 $J$=8--7 transition (e.\,g. Pardo et al. \cite{pardo}). In
 contrast, the $^{29}$SiO and $^{30}$SiO circumstellar emission has
 been less studied. In fact, maser amplification from these
 isotopomers has been measured in a few sources and a few rotational
 transitions. The first detection of a $^{29}$SiO maser line, the
 $v$=0 \mbox{$J$=1--0,} in a variable star was performed by Cho et al.
 (\cite{cho}).  Subsequent single-dish studies of the $v$\,$\geq$\,0
 \mbox{$J$=1--0} $^{29}$SiO circumstellar masers revealed that this
 rare isotopomer emission also had some properties usually associated
 to the $^{28}$SiO lines, such as the correlation with the \mbox{IR 8
   $\mu$m} radiation or its time variability (Alcolea \& Bujarrabal
 \cite{alcolea}).
 
 Our theoretical knowledge of these rare isotopomer emissions is still
 poor.  Some pumping mechanisms have been proposed, but they do not
 compare well with observations.  In order to constrain these models,
 we have studied several SiO maser lines in the OH/IR variable
 IRC\,+10011 (WX Psc) by means of Very Long Baseline
   Interferometry (VLBI) techniques.  We present in this paper
 interferometric maps for the 7\,mm $v$=1, $v$=2 $J$=1--0 and 3\,mm
 $v$=1 $J$=2--1 transitions of $^{28}$SiO, as well as the $v$=0
 \mbox{$J$=1--0} line of the $^{29}$SiO isotopomer.  For the first
 time, we have been able to detect and map the $^{29}$SiO maser line
 at high spatial resolution (better than 1 mas). We focus on
 the comparison between the maps of the different maser transitions
 and discuss how these results may affect the overall theoretical SiO
 pumping scenarios.

 \section{Observations and data analysis}
 
 Using the NRAO\footnote{The National Radio Astronomy Observatory is a
 facility of the National Science Foundation operated under
 cooperative agreement by Associated Universities, Inc.} Very Long
 Baseline Array (VLBA) we performed sub-milliarcsecond resolution
 observations of the SiO maser emission in the O-rich LPV star
 IRC\,+10011 on 2002 December 7. 
 \begin{figure}[!ht]
 \centering \includegraphics[angle=0, width=0.4\textwidth]{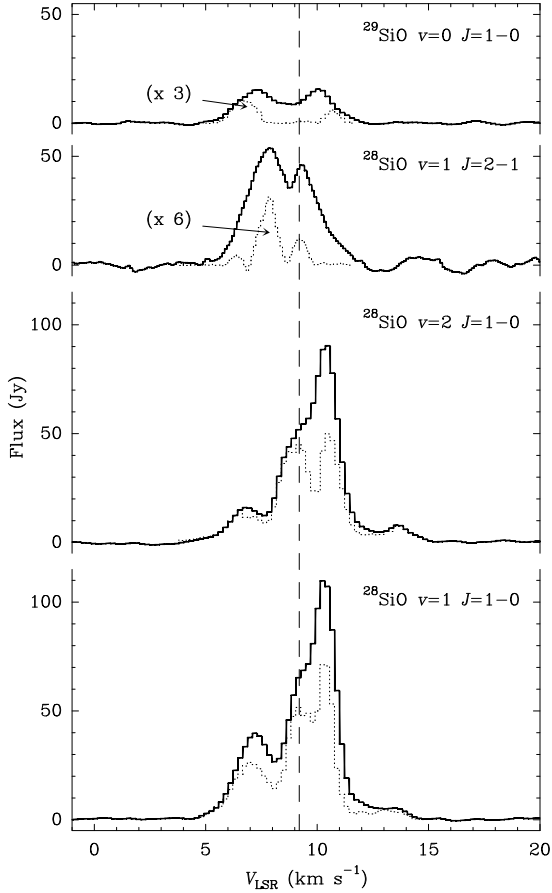}
 \caption{The total power (continuous lines) and recovered flux (dotted
   lines) spectra of the SiO masers observed in IRC\,+10011. The
   dashed line indicates the systemic velocity of the source
   ($V\rm{_{LSR}}$= 9.2 km\,s$^{-1}$, taken from Cernicharo et al.
   \cite{cerni2}).}
 \label{espec} 
 \end{figure}
 
 The results in this paper correspond to the second epoch of
 observations for IRC\,+10011, as part of a VLBA
 multi-epoch/transitional study of the SiO maser emission in AGB stars
 (see for further details Soria-Ruiz et al. \cite{soriaruiz}).
 The observed maser lines are summarized in Table \ref{tab1}.
 In this observational run all the SiO transitions were detected and
 mapped, including the $^{29}$SiO line. This is the first time that
 maser emission other than the main molecular species has been imaged
 using very long baseline interferometry.
 
 The data were correlated in Socorro (New Mexico).  In the 43 GHz
 observations, a bandwidth of 8 MHz was recorded and the correlator
 provided 256 frequency channels, thus achieving a spectral resolution
 of about 0.22 \,km\,s$^{-1}$.  For the 86 GHz transition, the
 bandwidth was 16 MHz and 512 spectral channels were used, being the
 resolution $\sim$\,0.11 \,km\,s$^{-1}$.  The calibration was done
 using the standard procedures for spectral line observations in the
 Astronomical Image Processing System (AIPS) package. For the final
 phase calibration we have used a strong maser component as reference,
 and therefore, all positions in the resulting maps are refered to the
 location of this spot, which is not the same for all the transitions.

 Fig. \ref{espec} presents for each of the SiO transitions, the total
 power spectrum of Los Alamos antenna, which was used as reference in
 the data reduction process, and the spectrum of the flux recovered
 in the maps.  
 The line profiles of the $v$=1 and $v$=2 $J$=1--0 masers are similar
 although they differ from the other two transitions.  About 75\% of
 the $v$=1 and $v$=2 $J$=1--0 maser emission was recovered after the
 calibration and imaging, while 10\% and 25\% of the emission was
 imaged for the $^{28}$SiO $v$=1 $J$=2--1 and $^{29}$SiO $v$=0
 $J$=1--0 respectively.

\setbox0=\hbox{$--$}

\begin{table}[t]
\setbox1=\hbox{0}
\setbox2=\hbox{$-$}
\caption{Summary of observational results}
\centering
\begin{tabular}{@{}cc|c@{\,\,\,}c@{\,\,\,}c@{}}
\cline{3-1}\cline{4-1}
\hline
\hline
&&&\\[-9pt]
maser&$\nu_{\rm{rest}}$&$R\rm{_{in}}$\,\,$R\rm{_{out}}$&\,restoring beam\\
transition & (MHz)  &(mas)&\hspace{\wd2} size (mas) \hspace{\wd2} PA (\degr)\\[2pt]
\hline\hline
&&&\\[-9pt]
 $^{28}$SiO\,\, $v$=1 $J$=1--0 &43122.080 &10.1\,\,13.1&\, 0.80$\times$0.53\,\,\, $-$15.6\\
\hspace{\wd0}\hspace{\wd0}$v$=2 $J$=1--0 &42820.587 &\hspace{\wd1}8.8\,\,12.3 &\, 0.85$\times$0.35\,\,\, $-$12.3\\
 $^{29}$SiO\,\, $v$=0 $J$=1--0 &42879.916 &11.8\,\,15.3&\, 0.97$\times$0.73\,\,\, \hspace{\wd1}$-$7.7  \\
$^{28}$SiO\,\, $v$=1 $J$=2--1 &86243.442 &14.7\,\,17.0&\, 0.58$\times$0.48\,\,\, \hspace{\wd2}16.1 \\[2pt]
\hline
\end{tabular}
\label{tab1}
\end{table}

\begin{figure*}[!th]

\vspace{8cm} 
  \includegraphics{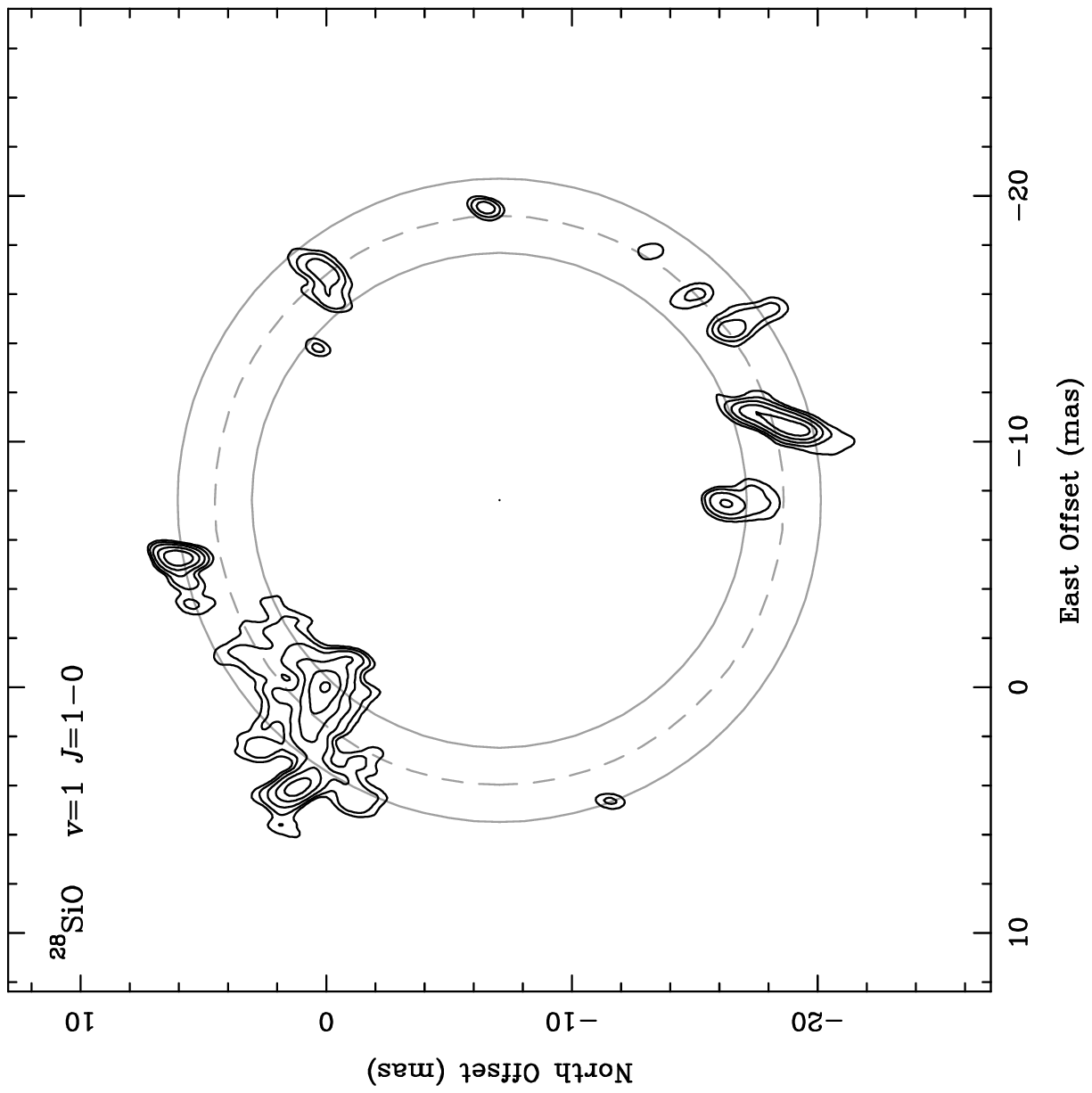} \includegraphics{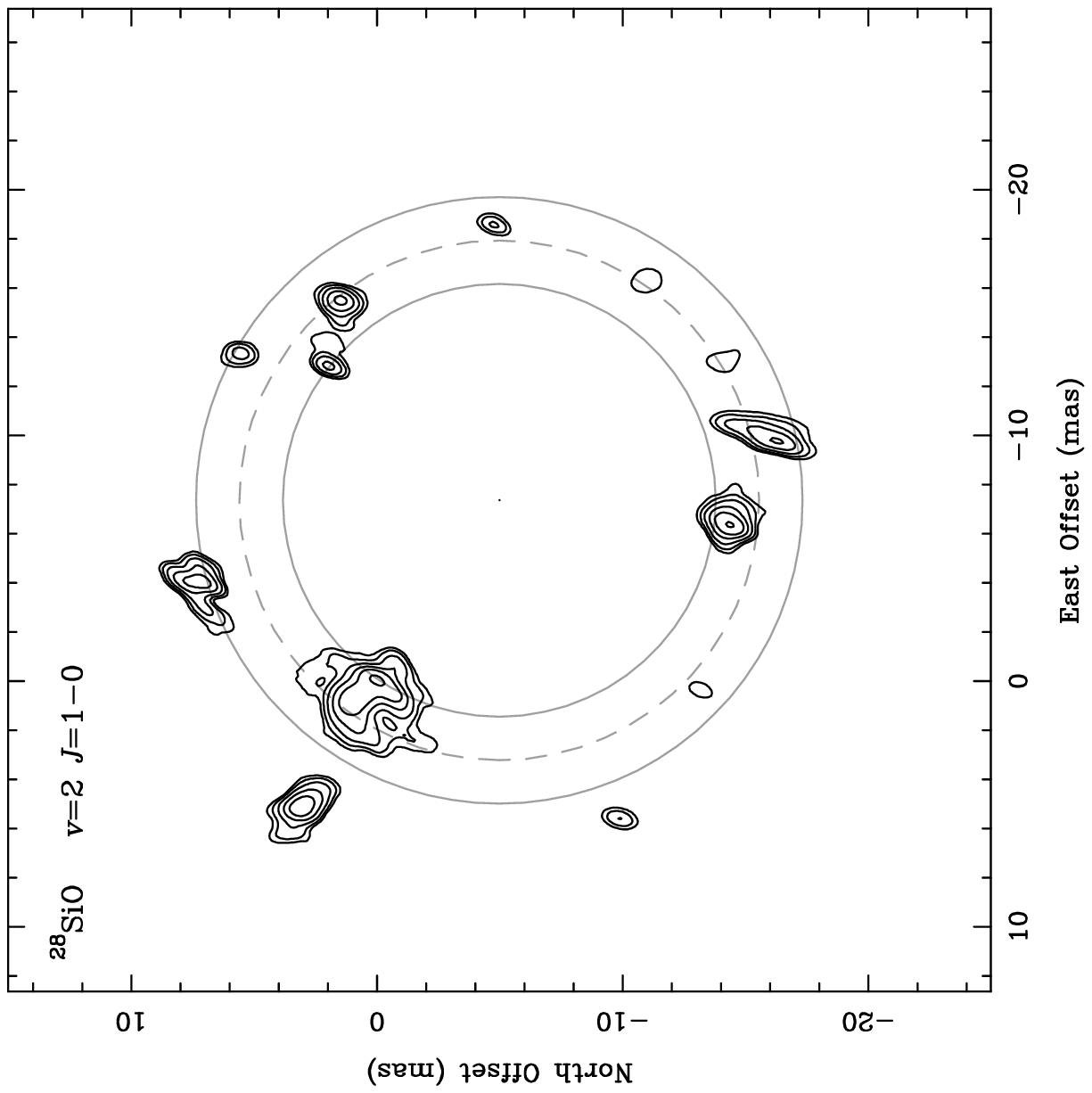}
  \vspace{8.3cm} 
  \includegraphics{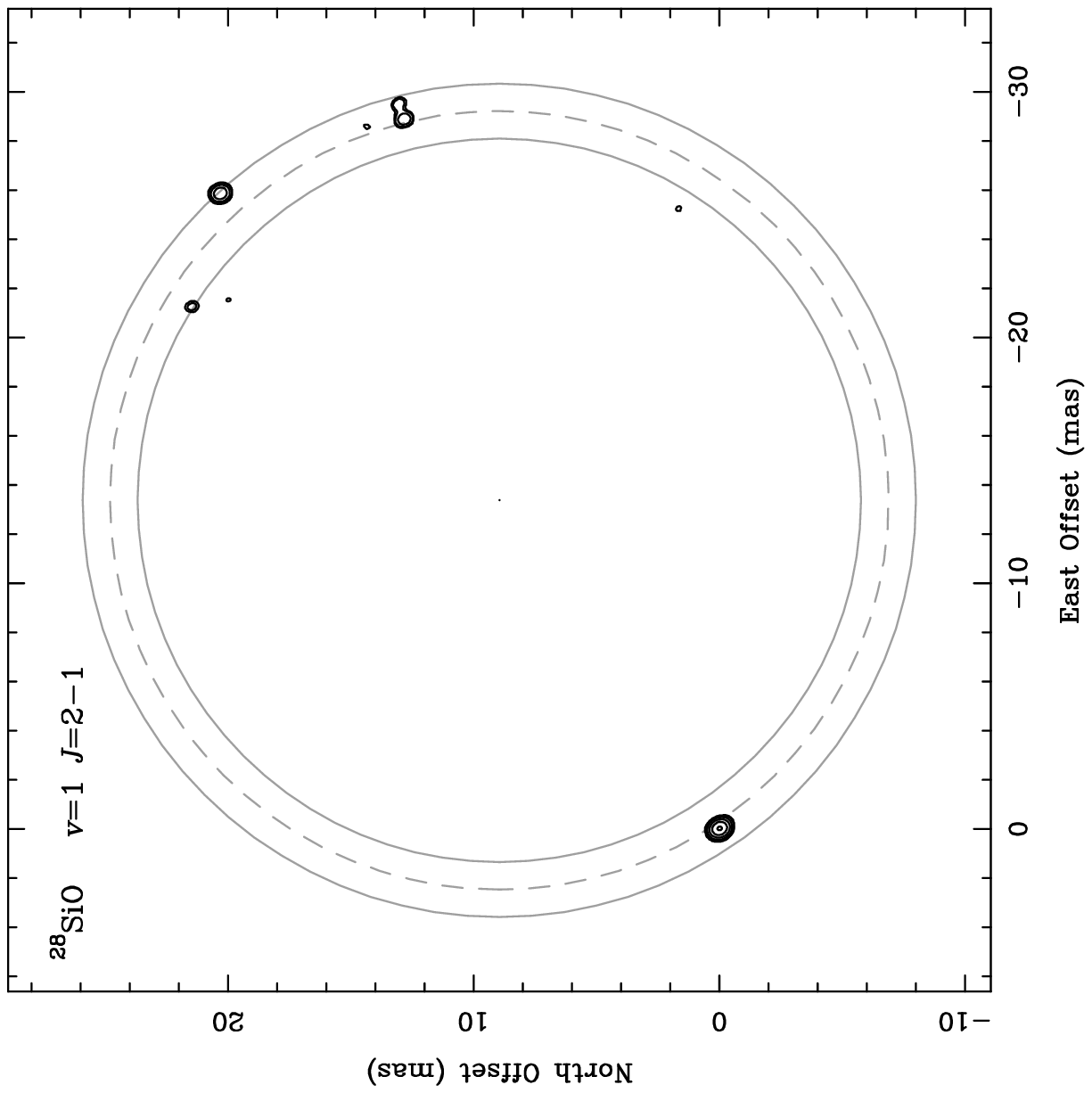} \includegraphics{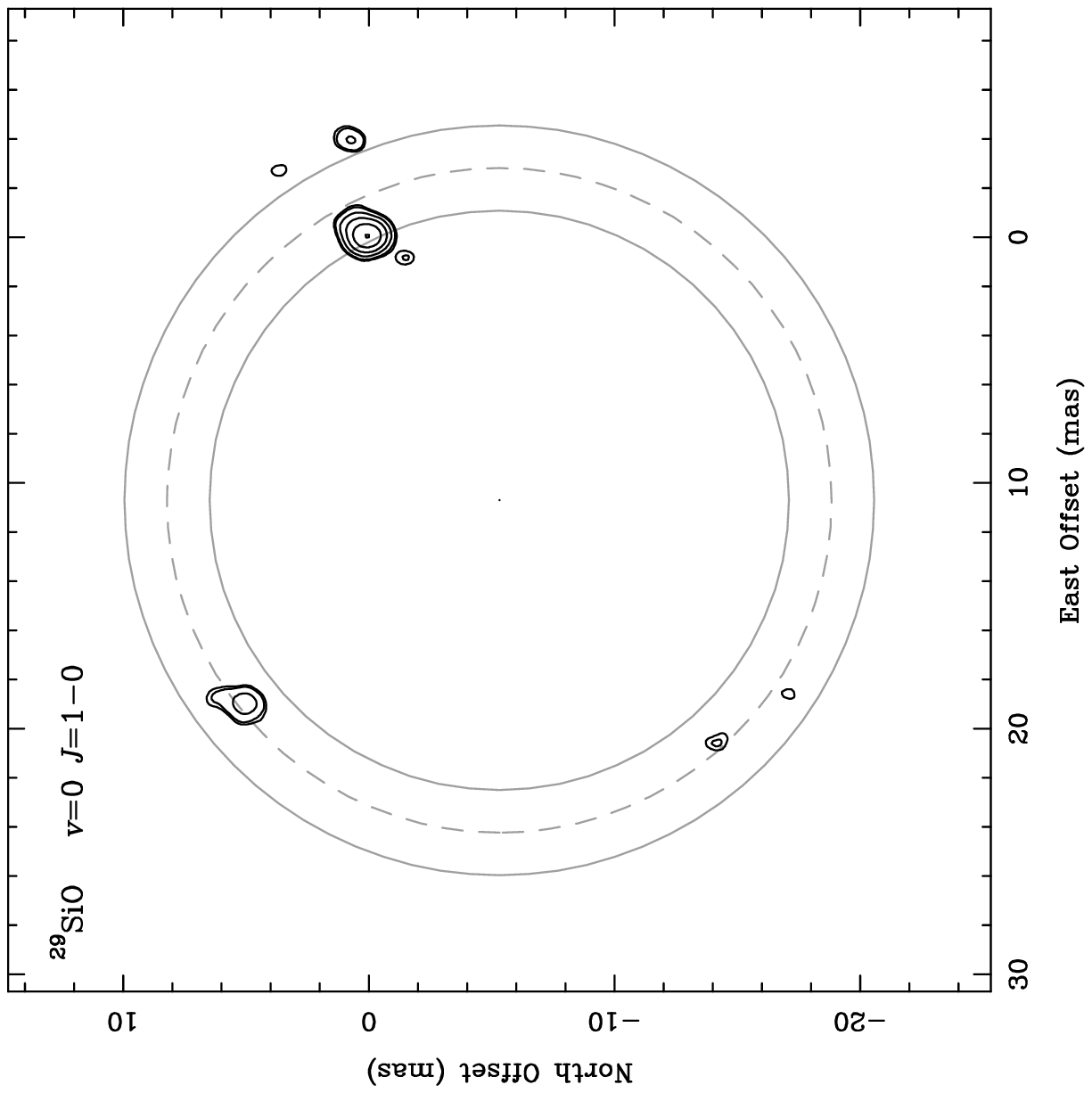}
\vspace{-0.1cm}
\caption{
  Integrated intensity maps for the SiO maser emissions in
  IRC\,+10011.  Top: $^{28}$SiO $v$=1 (left) and $v$=2 (right)
  $J$=1--0 rotational transitions. The peak flux is 9.14 and 9.72 Jy
  beam$^{-1}$ $\cdot$\,km\,s$^{-1}$ respectively, the contour levels
  in both images are 0.28, 0.57, 1.14, 2.28 4.16 and 8.34 Jy
  beam$^{-1}$ $\cdot$\,km\,s$^{-1}$ and the rms noise level 0.07 Jy
  beam$^{-1}$ $\cdot$\,km\,s$^{-1}$. Bottom: $^{28}$SiO $v$=1 $J$=2--1
  (left) and $^{29}$SiO $v$=0 $J$=1--0 (right) maser transitions.  The
  peak flux is 3.37 and 1.77 Jy beam$^{-1}$ $\cdot$\,km\,s$^{-1}$
  respectively.  The contour levels are 0.15, 0.20, 0.40,
  0.80, 1.60 and 3.20 Jy beam$^{-1}$ $\cdot$\,km\,s$^{-1}$ for the
  $J$=2--1 line, and 0.08, 0.12, 0.23, 0.46, 0.92 and 1.76 Jy
  beam$^{-1}\cdot$\,km\,s$^{-1}$ for the $J$=1--0 line. The rms noise
  levels in the maps are 0.02 and 0.01 Jy beam$^{-1}$
  $\cdot$\,km\,s$^{-1}$ respectively. Circles denote the ring fitting
  of the masing regions (dashed: mean radius $\bar{R}$, continuous:
  $R_{\rm{out}}$ and $R_{\rm{in}}$ defined as $\bar{R}\pm
  \frac{1}{2}\triangle{R}$).  We note that the four images have the
  same scale to ease the comparison between the emitting regions and
  that in all the cases the derived center of the ring is placed in
  the center of the box, while the origin of the map remains at the
  position of the reference spot (see Sect.\,2).}
\label{IRC} 
\end{figure*}

The integrated intensity maps are shown in Fig. \ref{IRC}.  As can be
seen in the upper panels, the spatial distributions of the $^{28}$SiO
$v$=1 and $v$=2 $J$=1--0 emissions appear to be similar, both are
ring-like, with the $v$=2 located in a slightly inner layer of the
envelope.  In contrast (see lower panels of Fig. \ref{IRC}), the
$^{28}$SiO $v$=1 $J$=2--1 and $^{29}$SiO $v$=0 \mbox{$J$=1--0} maser
maps are composed of a fewer number of spots and the ring distribution
is less clear.  In order to study and compare more accurately the
angular extent of the different masing regions, we fitted a ring to
our observational data.  To do this, we selected only the maser
features with a signal-to-noise ratio larger than 6 in at least three
consecutive spectral channels.  These calculations give the center,
the mean radius ($\bar{R}$) and the width of the ring ($\triangle{R}$,
being twice the standard deviation of the sample). We have summarized
in \mbox{Table \ref{tab1}} for the transitions observed, the inner and
outer radii derived from the fits (defined as $R_{\rm{in}} =
\bar{R}-\frac{1}{2}\triangle{R}$ and
$R_{\rm{out}}=\bar{R}+\frac{1}{2}\triangle{R}$) as well as the shape
of the gaussian restoring beam at half power (major and minor axis and
position angle of the major axis) of the resulting maps.

\section{Results}
 
 Previous long baseline interferometric studies of the SiO masers
 demonstrated the existence of very compact emission distributed in
 ring-like structures (see e.g. Diamond et al. \cite{diamond}
 and in IRC\,+10011 by Desmurs et al. \cite{desmurs}). Those
 results are limited to the $^{28}$SiO transitions. Therefore, our new
 detection and mapping of the $^{29}$SiO $v$=0 $J$=1--0 illustrates
 for the first time the spatial distribution of this rare
 isotopomer emission in evolved stars.
 
 The map of the $^{29}$SiO transition is composed of seven maser spots
 with velocities ranging from 5.7 to 11.3 km\,s$^{-1}$. The brightest
 feature has an intensity of \mbox{$\sim$1.77 Jy
 beam$^{-1}\cdot$\,km\,s$^{-1}$} (see bottom-right panel of
 \mbox{Fig. \ref{IRC})}.  This $^{29}$SiO emission is the weakest
 among the four lines observed, with an intensity 6 or 7 times lower
 than the $^{28}$SiO $J$=1--0 transitions. The emission forms a ring,
 though incomplete, with a mean radius of $\sim$13.5\,mas, therefore
 this maser radiation is probably amplified tangentially, as other
 well studied $^{28}$SiO maser transitions; in \mbox{Table
 \ref{tab1}} we show the inner and outer radii obtained from our
 maps following the method described in Sect.\,2.
 
 Both the $v$=1 and $v$=2 $J$=1--0 $^{28}$SiO emissions present a
 ring-like structure composed of 10--12 spots with a similar
 distribution.  In IRC\,+10011, Desmurs et al. (\cite{desmurs}) and
 Soria-Ruiz et al. (\cite{soriaruiz}) found a systematic shift between
 these two emitting regions of about 1--3 mas, with the $v$=2 radius
 being always smaller.  We confirm this result since the $v$=2
 ring radius is also $\sim$1 mas smaller than the $v$=1 one (see Table
 \ref{tab1} and Fig.  \ref{IRC}).  We note that a similar
 trend has also been observed in other Mira variables in several
 epochs over a stellar period (Cotton et al.  \cite{cotton}; Yi et
 al. \cite{yi}).
 From the three existing maps of these 7\,mm lines in IRC\,+10011
 (\mbox{Desmurs et al. \cite{desmurs},} \mbox{Soria-Ruiz et al.
   \cite{soriaruiz}} and this work, which correspond to phases,
 $\phi\sim$\,0.8, 0.1 and 0.5 respectively) changes of about 10\% and
 20\% are derived for the sizes of the $v$=1 and $v$=2 $J$=1--0 maser
 emitting regions respectively.
 
 From our maps of rotational transitions within the same vibrational
 state, that is, the $^{28}$SiO $v$=1 $J$=1--0 and $J$=2--1, we
 conclude that the latter maser is produced in a shell further away
 than the $J$=1--0 one (see Table \ref{tab1}).  Furthermore, the
 spatial distributions of the various components clearly differ
 \mbox{(Fig. \ref{IRC}).}  A similar relative location of the
 $^{28}$SiO $v$=1 $J$=1--0 and $J$=2--1 masing regions has been
 found in previous observations of IRC\,+10011 (Soria-Ruiz et al.
 \cite{soriaruiz}).  These authors have pointed out that there is a
 clear contradiction between the observational results, that is, the
 3\,mm transition in the $v$=1 being produced in an outer region of
 the envelope than the corresponding 7\,mm one, and the theoretical
 predictions, in which these two lines are spatially coincident 
 (either radiative or collisional models). The authors conclude that
 this discrepancy can be overcome when the overlap between
 ro-vibrational transitions of H$_{2}$O and $^{28}$SiO are
 introduced in the $^{28}$SiO excitation (see Soria-Ruiz et al.
 \cite{soriaruiz}).  Other comparisons of these 3 and 7\,mm lines
 have been done by Doeleman et al.  (\cite{doeleman}) in the Orion
 KL nebula, finding a similar result, and by Phillips et al.
 (\cite{phillips}) towards the AGB star R Cas.  In this case, the
 authors claim a similar distribution of both emissions.
 
 The $^{29}$SiO $v$=0 $J$=1--0 masing region appears in a layer
 located in between those of the $^{28}$SiO $v$=1 $J$=1--0 and the
 $^{28}$SiO $v$=1 $J$=2--1 emitting shells. This is in fact remarkable
 since these two $v$=1 transitions require excitation temperatures of
 about 2000 K and the $^{29}$SiO maser is a low-excitation line
 [$E$\,($J$=1)\,$\sim$\,2 K]. Therefore, the obtained location
 of this masing region suggests that the rotational levels involved in
 the inversion are populated mainly via de-excitations from upper
 $v$\,$>$\,0 states.

 \section{Discussion}
 
 Attempts to model the $^{29}$SiO maser amplification in evolved
 stars, and in particular the $v$=0 $J$=1--0 line, have been made
 using different excitation schemes.  Robinson \& Van Blerkom
 (\cite{robin}) and Deguchi \& Nguyen-Quang-Rieu (\cite{deguchi})
 proposed that the ground state $^{29}$SiO masers are produced if the
 vibrational transitions of this molecule present a significantly
 higher opacity along the radial direction than in the tangential one.
 This asymmetry with the direction is produced in slowly
 accelerated envelopes, yielding line profiles composed of two
 peaks separated by twice the expansion velocity of the envelope,
 similar to what is typically observed in circumstellar OH masers.
 These predictions are clearly in contradiction with the observations,
 since the line shapes of $^{29}$SiO masers are composed of narrow
 peaks near or at the stellar velocity (see also Alcolea \& Bujarrabal
 \cite{alcolea}).  Moreover, the map for IRC\,+10011 shows a ring-like
 geometry, very probably indicating tangential amplification (as it
 happens in the $^{28}$SiO maser lines). Finally, a small
 velocity gradient is not expected for the inner region of the
 envelope where the $^{29}$SiO maser is produced.
 
 The other proposed mechanism involves line overlaps between infrared
 transitions of $^{28}$SiO and $^{29}$SiO.  This effect was first
 suggested by Olofsson et al.  (\cite{olofsson}) to explain the $v$=0
 $J$=2--1 $^{29}$SiO maser.  Subsequent calculations by
 Gonz\'alez-Alfonso \& Cernicharo (\cite{gonzalez}) suggest that the
 $v$=0 $J$=1--0 $^{29}$SiO line can be efficiently pumped by the
 overlap between the $^{28}$SiO $v$=2--1 $J$=4--3 and the $^{29}$SiO
 \mbox{$v$=1--0 $J$=1--0} ro-vibrational transitions.  
 
 These models do not provide calculations of the size and brightness
 distribution of the emitting regions, thus making it very difficult
 to conclude on the compatibility of the theory with the presented
 VLBI observations.
 
 In any case, our observational results are in agreement with the
 predictions of the mechanism proposed by Gonz\'alez-Alfonso \&
 Cernicharo (\cite{gonzalez}).  For this model to work, the
 ro-vibrational transitions $v$=2--1 of $^{28}$SiO must have
 relatively high opacities ($\tau$$\gtrsim$\,1). This condition is
 also required for the arising of the strong $^{28}$SiO $v$=1 and 2
 masers (in both collisional and radiative excitation schemes).
 Therefore, the fact that the $^{29}$SiO emission arises in the same
 inner shell as the $^{28}$SiO $v$=1 and 2 masers guarantees that the
 main requirement of the model by Gonz\'alez-Alfonso \& Cernicharo
 (\cite{gonzalez}) is satisfied.  (We also note that the $^{29}$SiO
 $v$=1--0 $J$=1--0 line also coincides in frequency, better than 2 km
 s$^{-1}$, with the ro-vibrational line $\nu_{2}$=2\,--\,1
 7$_{2,6}$\,--\,8$_{3,5}$ of H$_{2}$O.)  Finally, as it has been
 mentioned in the previous section, overlaps also appear to be
 responsible for the discrepancies found between observations of some
 rotational lines of $^{28}$SiO and models (see also Bujarrabal et al.
 \cite{bujarrabal}, Herpin \& Baudry \cite{herpin}).

 We are currently carrying out similar high resolution studies of
 maser lines of $^{28}$SiO and $^{29}$SiO in other evolved stars, to
 see if our results are reproduced in other targets, and especially
 the observed location of the emitting region and spatial distribution
 of the $^{29}$SiO $v$=0 $J$=1--0 line.  A more extended sample will
 also contribute to a better understanding of which processes may be
 involved in the pumping and if the line overlaps play such an
 important role in the SiO excitation.

\begin{acknowledgements}
  This work has been financially supported by the Spanish DGI (MCYT)
  under projects AYA2000-0927 and AYA2003-7584.
\end{acknowledgements}

\end{document}